\documentclass[a4paper,11pt, draftcls, onecolumn]{IEEEtran}
\usepackage{graphics, graphicx}
\usepackage{amssymb,amsmath}
\usepackage{amsmath}
\usepackage[rm]{subfigure}   
\usepackage{threeparttable}  
\usepackage{color}
\usepackage{url}
\usepackage[T1]{fontenc}
\usepackage{algorithm}
\usepackage{cite}
\usepackage{multirow}
\usepackage[normalem]{ulem}
\usepackage{float}
\usepackage{flafter}
\usepackage{algorithmic}

\usepackage{xspace,colortbl}
\usepackage{tikz}
\usepackage{amsmath}

\usepackage{algorithm}
\usepackage{indentfirst}
\usepackage{setspace}

\newcommand\scalemath[2]{\scalebox{#1}{\mbox{\ensuremath{\displaystyle #2}}}}

\begin{document}
\title{A Grant-based Random Access Protocol in Extra-Large Massive MIMO System}
\author{Ot\'avio Seidi Nishimura, Jos\'e Carlos Marinello Filho, Taufik Abr\~ao
\noindent\thanks{Copyright (c) 2015 IEEE. Personal use of this material is permitted.} 
\noindent\thanks{O. S. Nishimura, T. Abr\~ao are with Electrical Engineering Department, State University of Londrina, PR, Brazil.}
\noindent\thanks{J. C. Marinello is with Electrical Engineering Department, Federal University of Technology PR, Corn\'elio Proc\'opio, PR, Brazil.  ({nishimuraotavio@gmail.com, jcmarinello@utfpr.edu.br, taufik@uel.br}).}
\thanks{This work was supported in part by the  Coordination for the Improvement of Higher Education Personnel (CAPES)-Brazil, Finance Code 001, by the Arrangement between the European Commission (ERC) and the Brazilian National Council of State Funding Agencies (CONFAP), CONFAP-ERC Agreement H2020, by the National Council for Scientific and Technological Development (CNPq) of Brazil under grants 404079/2016-4 and 310681/2019-7.}
}

\maketitle

\begin{abstract}
Extra-large massive multiple-input multiple-output (XL-MIMO) systems is a new concept, where spatial non-stationarities allow activate a high number of user equipments (UEs). This paper focuses on a grant-based random access (RA) approach in the novel XL-MIMO channel scenarios. Modifications in the classical Strongest User Collision Resolution (SUCRe) protocol have been aggregated to explore the visibility regions (VRs) overlapping in XL-MIMO. The proposed grant-based RA protocol takes advantage of this new degree of freedom for improving the number of access attempts and accepted UEs. As a result, the proposed grant-based protocol for XL-MIMO systems is capable of reducing latency in the pilot allocation step.
\end{abstract}
\begin{IEEEkeywords}
Random access protocol, Grant-based, massive MIMO, XL-MIMO, non-stationarity, visibility region (VR).
\end{IEEEkeywords}

\section{Introduction}\label{sec:intro}
As stated by the METIS (mobile enablers twenty-twenty society) project \cite{metis}, there is a predicted rapidly increase in the demand of network access and data traffic for the next few years coming. To enable such requirement the fifth generation of wireless networks (5G) is expected to provide three main services: enhanced Mobile Broadband (eMBB), Ultra Reliable Low-Latency Communication (URLLC) and massive Machine Type Communication (mMTC). Another awaited scenario is crowded Mobile Broadband (cMMB), where the number of UEs surpasses those of available pilot sequences and very high data rate is demanding. 

Channel state information (CSI) is necessary to provide coherent communication and this is implemented by using orthogonal pilots. However, the number of {UEs} in crowded scenarios is much greater than the available pilot sequences, causing an unfeasible situation to schedule. There are different methods of RA, which can be classified in two types: random access to pilots (RAP) and random access to pilots and data transmission (RAPiD) \cite{randomAccess2017}. The second approach is a grant-free RA and uses pilot hopping in multiple time slot transmissions, managing pilot collisions and interference with massive MIMO (mMIMO) properties \cite{erapid}, \cite{CPA}. 

This paper focuses in RAP, a grant-based RA; herein the transmissions happen in an RA pilot domain and several UEs are trying to acquire a dedicated pilot for a collision free connection. A promising protocol to handle many sporadic access attempts is the SUCRe \cite{sucre2017}. In general, it resolves RA pilot collisions, in a totally distributed way, choosing the strongest colliding user and it is well settled in a crowded mMIMO system.

Since mMIMO is already an essential enabler for 5G networks, in \cite{reality2019mMIMO} five challenges for this technique have been discussed. One of them is to establish how the several conventional {mMIMO} approaches will be structured in extra large arrays. These arrays can be implemented under several types of infrastructures, as buildings, stadiums, or shopping malls, where UEs are mainly placed near the panels generating non-stationary VRs.

The paper {\it contribution} consists in proposing a grant-based RA protocol to operate advantageously in XL-MIMO systems, in which the large array size and the proximity with the users give rise to spatial non-stationarities across the array. In such configuration, it is possible to take advantage of UEs distinct VRs as an additional degree of freedom in order to improve the system performance while reducing the latency in the pilot allocation step.

\noindent{\it Notation}: The conjugate, transpose and conjugate-transpose of a matrix $\mathbf{A}$ are represented by $\mathbf{A}^*$, $\mathbf{A}^T$ and $\mathbf{A}^H$, respectively. $\mathbf{I}_M$ is the $M\times M$ identity matrix, $\left| \cdot \right|$ and $\left\| \cdot \right\|$ represent the cardinality of a set and the Euclidean norm of a vector, respectively. Operators $\mathbb{E}\{\cdot\}$, and $\mathbb{V}\{\cdot\}$ denote the expectation and the variance of a random variable. $\mathcal{N}(.,.)$ denotes a Gaussian distribution, $\mathcal{CN}(.,.)$ represents a circularly-symmetric
complex Gaussian distribution, and $\mathcal{B}(.,.)$ represents a binomial distribution. $\mathbb{C}$ and $\mathbb{R}$ denote spaces of complex and real-valued numbers, while $\Gamma(\cdot)$ represents a Gamma function. The operator that gives the real part of its argument is $\Re(.)$

\section{System model}\label{sec:model}
\begin{figure}[htbp!]
	\centering
	\includegraphics[width=.66\linewidth]{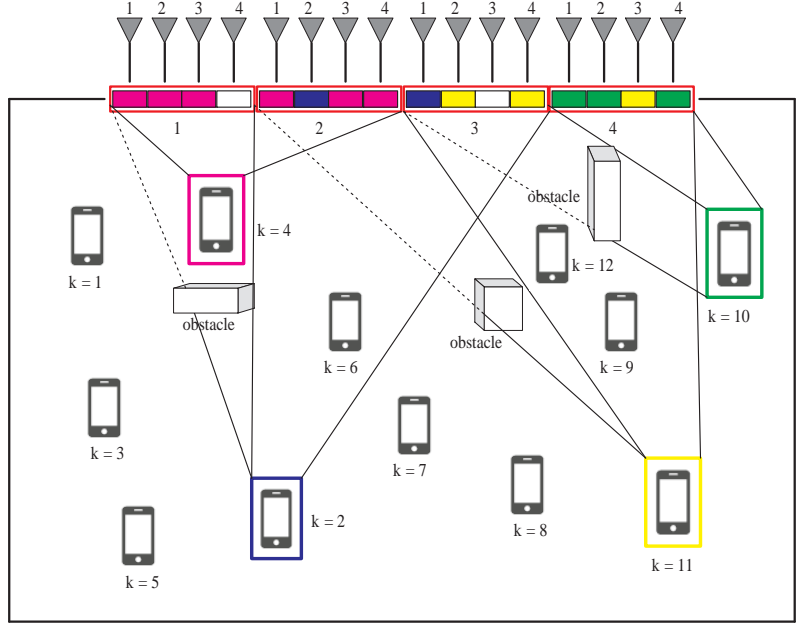}
	\vspace{-4mm}
	\caption{Example of a uniform linear extra large array with $B = 4$ SAs, each with $M_b = 4$ antennas. UEs have different VRs and consequently distinct associated SAs and channel gains to establish communication. There are $K = 12$ iUEs, but only UEs $k = 2,4,10,11$ want to become active.}
	\vspace{-4mm}
	\label{fig:array}
\end{figure}

For simplicity,  our adopted XL-array is a uniform linear array (ULA, Fig. \ref{fig:array}), operating in time-division-duplexing (TDD). Since channel modeling is not the focus of this work, it is assumed a simplified bipartite graph model in XL-MIMO, as the one used in \cite{amiri2018}. Accordingly, the system is divided into $B$ subarrays (SAs), each composed by a fixed number of $M_b = M/B$ antennas.  
Let $\mathcal{M}$ be the set composed by $1,..,B$, and $\mathcal{V}_k \subset \mathcal{M}$ be the subset of visible SAs associated to user $k$. To model the VR set $\mathcal{V}_k$ at random, each SA is independent and identically distributed (i.i.d.) following a Bernoulli distribution with success probability $P_b$. Then, every UE has a binary vector of size $B$ to indicate if each SA is visible (1) or not (0). For simulation purposes, $|\mathcal{V}_k|>0, \; \forall \; k$.  

 Let $\mathcal{K} = \mathcal{U}\backslash\mathcal{A}$ be the set of inactive UEs (iUEs), where $\mathcal{U}$ is the set of UEs in the entire cell, and $\mathcal{A}	\subset \mathcal{U}$ is the subset of active users, each with their dedicated payload pilot. Thus, $K = |\mathcal{K}|$ represents the number of {iUEs}. Let $\tau_p$ denote the number of mutually orthogonal pilot sequences $\boldsymbol{s}_1, ...,\boldsymbol{s}_{\tau_p}\in \mathbb{C}^{\tau_p \times 1}$. In this case, each pilot has length $\tau_p$ and $\lVert \boldsymbol{s}_t \rVert^2 = \tau_p$.

In this work, it is considered a sliced channel vector $\mathbf{h}^{(b)}_k \in \mathbb{C}^{M_b\times 1}$ between UE $k\in \mathcal{K}$ and the $b$-th {SA} with $M_b$ antennas. The vector follows a Rayleigh fading channel model
\begin{equation}\label{eq: uncorr Rayleigh channel}
\mathbf{h}^{(b)}_{k}\sim \mathcal{CN}(0,\beta^{(b)}_k\mathbf{R}^{(b)}_{k}),
\end{equation}
for all users $k=1,2,...,K$, each with a large scale fading coefficient $\beta^{(b)}_k$. When assuming i.i.d. fading channel, $\mathbf{R}^{(b)}_{k} = \mathbf{I}_{M_b}$, while for correlated fading channels we have \begin{equation}\label{eq:Rcorrel}
[\mathbf{R}^{(b)}_{k}]_{i,\ell} = r^{-|\ell-i|} e^{j\theta^{(b)}_k (\ell-i)},  
\end{equation}
where $\theta^{(b)}_k$ is the angle between $k$-th UE and the $b$-th {SA}, and $r\in (0;1)$ is the correlation index.
Actually, a {UE} has one coefficient per antenna, since the BS is an extra large array. To simplify, $\beta^{(b)}_k$ assumes the mean value considering all antennas of {SA} $b$, $\beta^{(b)}_k = \frac{1}{M_b}\sum_{m =1}^{M_b}\beta^{(b)}_{k,m}$, where $\beta^{(b)}_{k,m}$ is the coefficient between UE $k$ and antenna $m$ ($m = 1,...,M_b$) at the $b$-th {SA}. In addition, invisible {SAs} for the $k$th {UE}, $b \notin \mathcal{V}_k$, are assumed to have $\beta^{(b)}_k = 0$. Moreover, herein, a urban micro scenario model \cite{3gpp.25.996} is considered:
\begin{equation}
 \beta^{(b)}_{k,m} = 10^{-\kappa \log(d^{(b)}_{k,m}) + \frac{g+\varphi}{10}},
\end{equation}
where $d^{(b)}_{k,m}$ represents the distance between UE $k$ and antenna $m$ ($m = 1,...,M_b$) at the $b$-th {SA}, $g = - 34.53$ dB is the pathloss at the reference distance, the pathloss exponent $\kappa = 3.8$, and $\varphi \sim \mathcal{N}(0,\sigma^2_{\rm sf})$  is the shadow fading, a log-normal random variable with standard deviation $\sigma_{\rm sf} = 10$ dB. 

Each {iUE} realizes a RA attempt with probability $P_a\leq 1$. User $k \in \mathcal{K}$ uniformly selects an uplink RA pilot sequence $\boldsymbol{s}_{r(k)} \in \mathbb{C}^{\tau_p\times 1}$, where $r(k)\in \{1,2,...,\tau_p \}$. Since transmission is uncoordinated, it is possible and usual that more than one UE choose the same pilot sequence $\boldsymbol{s}_t$. Therefore, let $\mathcal{S}_t = \{k:r(k) =t, \rho_k >0\}$ represent the set of {iUEs} indices transmitting pilot $t$, with power $\rho_k$. The cardinality of this set follows a binomial distribution \cite{sucre2017}:
\begin{equation}\label{eq: binomial distribution}
|\mathcal{S}_t|\sim \mathcal{B}\left(K,\frac{P_a}{\tau_p}\right).
\end{equation}

\begin{figure}[htbp!]
\vspace{-2mm}
\centering
\includegraphics[width=0.75\linewidth]{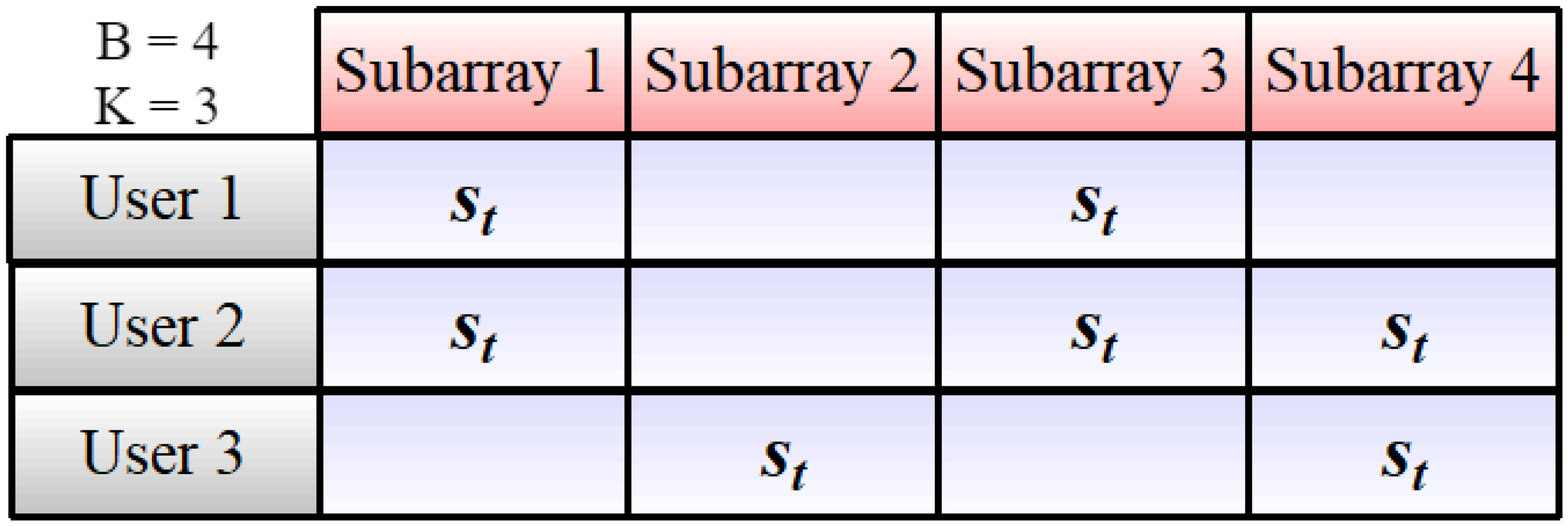}
\vspace{-3mm}
\caption{An example of the proposed UL arrangement with a probability $P_a = 1$, $K=3$ users, $B=4$ subarrays and $\tau_p=1$ available pilot sequence.}	
\label{fig:pilots}
\end{figure}

Fig. \ref{fig:pilots} depicts an arbitrary uplink RA arrangement with $K=3$, $B = 4$, $\tau_p = 1$ and $P_a = 1$. In this case, there are collisions in {SAs} 1, 3 and 4 between users 1 and 2, and 2 and 3, but no collisions between users 1 and 3. 

SUCRe protocol relies on {mMIMO} properties, as channel hardening and asymptotic favorable propagation:
\begin{eqnarray}
\frac{\lVert \mathbf{h}_k^{(b)} \rVert^2}{M_b}\xrightarrow{M_b\rightarrow\infty} \beta^{(b)}_k, \quad \forall k,b\label{eq: channel Hardening 1} \\
\frac{\mathbf{h}_k^{(b)H} \mathbf{h}_{k'}^{(b')}}{M_b} \xrightarrow{M_b\rightarrow\infty} 0, \quad \forall (k,b)\neq (k',b'), 
\label{eq: favorable propagation between UEs}
\end{eqnarray}
respectively. From eq. \eqref{eq: channel Hardening 1}, it follows that
\begin{eqnarray}
\sum_{j\in \mathcal{V}_k}\frac{\lVert \mathbf{h}_k^{(j)} \rVert^2}{M_b}\xrightarrow{M_b\rightarrow\infty} \sum_{j\in \mathcal{V}_k}\beta^{(j)}_k, \quad \forall k\label{eq: channel Hardening sum 2} \end{eqnarray}
which represents the overall channel gain over the visible {SAs} for $k$-th UE. Notice that the number of antennas per {SA}, $M_b$, does not always remain large, since VRs represent just a portion of antennas available for each user in a specific time. Nevertheless, the proposed protocol, {named SUCRe-XL,} still presents a satisfying performance even under certain reduced number of antennas per {SA}. 

\section{Proposed SUCRe-XL protocol}
We first describe how a straightforward adaptation of the conventional SUCRe protocol to the XL-MIMO scenario would be, demonstrating why it does not work. Then, we propose the SUCRe-XL protocol deploying modifications to operate in the XL-MIMO regime in step 2. The section concludes with the definition of the contention resolution rules and allocation strategy for the dedicated payload pilots.

\noindent\textit{\textbf{Step 1}: Random UL Pilot Sequence}. All UEs that want to be active send RA pilot sequences. In the BS, the $b$th SA receives signal $\textbf{Y}^{(b)}\in \mathbb{C}^{M_b\times\tau_p}$:
\begin{equation}\label{eq:step1: UE RA pilot}
\textbf{Y}^{(b)} = \sum_{k\in \mathcal{K}} \sqrt{\rho_k}\mathbf{h}_k^{(b)} \boldsymbol{s}_{r(k)}^{T} + \textbf{N}^{(b)},
\end{equation}
where $\textbf{N}^{(b)}\in \mathbb{C}^{M_b\times \tau_p}$ is the receiver noise, with i.i.d. elements distributed as $\mathcal{CN}(0,\sigma^2)$. To estimate the channel of UEs $k \in \mathcal{S}_t$ ($t = 1,...,\tau_p$), the BS correlates $\textbf{Y}^{(b)}$ for each sub-array $b$ with each normalized pilot sequence $\boldsymbol{s}_t$, 
\begin{equation}\label{eq:step1: Correlated RA pilot at the BS}
\textbf{y}_t^{(b)} = \textbf{Y}^{(b)}\frac{\boldsymbol{s}_t^*}{\lVert \boldsymbol{s}_t \rVert} = \sum_{i\in \mathcal{S}_t} \sqrt{\rho_i \tau_p}\mathbf{h}_i^{(b)}  + \mathbf{n}_t, \quad b=1,\ldots, B.
\end{equation} 
where $\mathbf{n}_t = \mathbf{N}\frac{\boldsymbol{s}_t^*}{\lVert \boldsymbol{s}_t \rVert} \sim \mathcal{CN}(0,\sigma^2 \mathbf{I}_{M_b})$ is the effective receiver noise. With eq. \eqref{eq: channel Hardening 1} and \eqref{eq: favorable propagation between UEs}, the following approximation holds: 

\vspace{-4mm}
\begin{equation}\label{eq:step1: asymptoticPropagation of y}
\frac{\lVert \sum_{b\in \mathcal{M}}\mathbf{y}^{(b)}_t \rVert^2}{M_b} \xrightarrow{M_b\rightarrow\infty} \underbrace{\sum_{b\in \mathcal{M}} \sum_{i \in \mathcal{S}_t} \rho_i \beta^{(b)}_i \tau_p}_{\alpha_{t}} \,\, +\,\, B\sigma^2.
\end{equation}
The proof of property (\ref{eq:step1: asymptoticPropagation of y}) is found in the appendix. Hence, the {\it sum of the signal gains}, $\alpha_{t}$, received at the BS for each RA pilot in step 1 is readily identified as the first term in \eqref{eq:step1: asymptoticPropagation of y}.

%
\noindent\textit{\textbf{Step 2}: Precoded Random Access DL Response}. In the second step of the SUCRe procedure, each SA responds with an orthogonal precoded DL pilot $\mathbf{V}^{(b)}\in \mathbb{C}^{M_b\times \tau_p}$. Using a normalized conjugate of $\mathbf{y}^{(b)}_t$, results:
\begin{equation}\label{eq:step2: DL pilot response2}
\mathbf{V}^{(b)} = \sqrt{\frac{q}{B}}\sum_{t = 1}^{\tau_p}\frac{\mathbf{y}_t^{(b)*}}{\lVert \mathbf{y}^{(b)}_t \rVert}\boldsymbol{s}_t^H, \quad b=1,\ldots,B,
\end{equation}
where $q$ is the predefined DL transmit power. 
Then, UE $k \in \mathcal{S}_t$ receives signal $\mathbf{v}_k^T \in \mathbb{C}^{1 \times \tau_p}$ given by
\begin{equation}\label{eq:step2:rx_signal_UE 0}
\textbf{v}_k^T = \sum_{m \in \mathcal{V}_k}\mathbf{h}_k^{(m)T} \mathbf{V}^{(m)} + \boldsymbol{\eta}_k^T,
\end{equation}
where  $\boldsymbol{\eta}_k \sim \mathcal{CN}(0,\sigma^2 \textbf{I}_{\tau_p})$ is the receiver noise. Next, each UE correlates the received signal in eq. \eqref{eq:step2:rx_signal_UE 0} with RA pilot $\boldsymbol{s}_t$:
\begin{equation}\label{eq:step2: correlated received signal at the UE 0}
v_k = \mathbf{v}_k^T \frac{\boldsymbol{s}_t}{\lVert \boldsymbol{s}_{t}  \rVert} = \sqrt{\frac{q \tau_p}{B} } \sum_{m\in \mathcal{V}_k} \mathbf{h}_k^{(m)T} \frac{\mathbf{y}_t^{(m)*}}{\lVert \mathbf{y}^{(m)}_t \rVert}  + \eta_k,
\end{equation}
where $\eta_k = \boldsymbol{\eta}_k^T  \frac{\boldsymbol{s}_t}{\lVert \boldsymbol{s}_t  \rVert} \sim \mathcal{CN}(0,\sigma^2)$ is the effective receiver noise. Dividing the equation by $\sqrt{M_b}$, and considering that asymptotic conditions of eq. \eqref{eq: channel Hardening 1} and \eqref{eq: favorable propagation between UEs} hold, it follows that: 
\begin{align}\label{eq:step2: zk-assymptoticFavorablePropagation 0}
\frac{v_k}{\sqrt{M_b}} = \sqrt{\frac{q \tau_p}{B}}\sum\limits_{m\in \mathcal{V}_k}\frac{\left(\mathbf{h}_k^{(m)H} \mathbf{y}^{(m)}_t \right)^*}{M_b \sqrt{\frac{1}{M_b} \left\lVert \mathbf{y}^{(m)}_t \right\rVert^2}}  + \frac{\eta_k}{\sqrt{M_b}} \\\nonumber
\xrightarrow{M_b\rightarrow \infty} \sum\limits_{m\in \mathcal{V}_k} \frac{\sqrt{\rho_k q /B  }\tau_p \beta^{(m)}_k }{\sqrt{ \sum\limits_{i \in\mathcal{S}_t} \rho_i \beta^{(m)}_i \tau_p + \sigma^2}}.
\end{align}
Notice that the magnitude $\alpha_{t}$ received at the BS, as in eq. \eqref{eq:step1: asymptoticPropagation of y}, cannot be mathematically separated, due to the sum of different denominators. Since the users cannot obtain this information, the application of the \emph{strongest user} criterion becomes difficult. For this reason, the following SUCRe for XL-MIMO protocol is proposed.\\
\vspace{-4mm}

\noindent\textit{SUCRe-XL Precoded DL Response}. In the second step of the SUCRe-XL protocol, instead of employing conjugate-$\mathbf{y}^{(b)}_t$ precoding  as in eq. \eqref{eq:step2: DL pilot response2}, all SAs use the same precoding vector $\sum_{b\in \mathcal{M}}\mathbf{y}^{(b)}_t$. Thus, each SA responds with the same signal $\mathbf{V}_{\textsc{xl}}\in \mathbb{C}^{M_b\times \tau_p}$:
\begin{equation}\label{eq:step2: DL pilot response}
\mathbf{V}_{\textsc{xl}} = \sqrt{\frac{q}{B}}\sum_{t = 1}^{\tau_p}\frac{\sum_{b\in \mathcal{M}}\mathbf{y}_t^{(b)*}}{\lVert \sum_{b\in \mathcal{M}}\mathbf{y}^{(b)}_t \rVert}\boldsymbol{s}_t^H.
\end{equation}
Then, the UE $k \in \mathcal{S}_t$ receives signal $\mathbf{z}_k^T \in \mathbb{C}^{1 \times \tau_p}$,
\begin{equation}\label{eq:step2:rx_signal_UE}
\textbf{z}_k^T = \sum_{m \in \mathcal{V}_k}\mathbf{h}_k^{(m)T} \mathbf{V}_{\textsc{xl}} + \boldsymbol{\eta}_k^T,
\end{equation}
and correlates it with RA pilot $\boldsymbol{s}_t$:

\begin{equation}\label{eq:step2: correlated received signal at the UE}
z_k = \mathbf{z}_k^T \frac{\boldsymbol{s}_t}{\lVert \boldsymbol{s}_{t}  \rVert} = \sqrt{\frac{q \tau_p}{B}} \sum_{m\in \mathcal{V}_k} \mathbf{h}_k^{(m)T} \frac{\sum_{b\in \mathcal{M}} \mathbf{y}_t^{(b)*}}{\lVert \sum_{b\in \mathcal{M}}\mathbf{y}^{(b)}_t \rVert}  + \eta_k.
\end{equation}
\vspace{-1mm}
In the same way of eq. \eqref{eq:step2: zk-assymptoticFavorablePropagation 0}, it follows that:
 \begin{align}\label{eq:step2: zk-assymptoticFavorablePropagation}\nonumber
\frac{z_k}{\sqrt{M_b}} = \sqrt{\frac{q \tau_p}{B}}\frac{\left(\sum\limits_{m\in \mathcal{V}_k}\mathbf{h}_k^{(m)H} \sum\limits_{b\in \mathcal{M}}\mathbf{y}^{(b)}_t \right)^*}{M_b \sqrt{\frac{1}{M_b} \left\lVert \sum\limits_{b\in \mathcal{M}} \mathbf{y}^{(b)}_t \right\rVert^2}}  + \frac{\eta_k}{\sqrt{M_b}} \\\nonumber
\xrightarrow{M_b\rightarrow \infty}\frac{\sqrt{\rho_k q /B}\tau_p \sum\limits_{m\in \mathcal{V}_k} \beta^{(m)}_k }{\sqrt{ \sum\limits_{b\in \mathcal{M}} \sum\limits_{i \in\mathcal{S}_t} \rho_i \beta^{(b)}_i \tau_p + B\sigma^2}}.
\end{align}
Thus, noise and estimation errors in the imaginary part are removed from eq. (\ref{eq:step2: correlated received signal at the UE}), resulting 
\begin{equation}
\frac{\Re(z_k)}{\sqrt{M_b}} \approx \frac{\sqrt{\rho_k q /B} \sum_{m\in \mathcal{V}_k} \beta^{(m)}_k \tau_p}{\sqrt{\alpha_{t} + B\sigma^2}}.
\end{equation}
Hence, the  $k$th UE can now have an estimate by isolating $\alpha_{t}$. The estimator of \cite{sucre2017} can be readily adapted to our RA XL-MIMO scenario as
\begin{align}\label{eq: a Approx2}
\widehat{\alpha}_{t,k} = &\max\left[\rho_k \sum_{m\in \mathcal{V}_k} \beta^{(m)}_k \tau_p \,\, , \right. \\ 
& \left. \left(\frac{\Gamma(M_b+1/2)}{\Gamma(M_b)}\right)^2 \frac{\rho_k  q \tau_p^2  \left(\sum_{m\in \mathcal{V}_k} \beta^{(m)}_k\right)^2}{B[\mathbb{\Re}(z_k)]^2}  - B\sigma^2 \right].\nonumber
\end{align}
\vspace{-3mm}

It is proved that changing the precoding as in eq. \eqref{eq:step2: DL pilot response} and adapting the $\widehat{\alpha}_{t,k}$ estimator as in eq. \eqref{eq: a Approx2} are sufficient to implement the proposed RA protocol in XL-MIMO scenarios. Such procedure does not cause any additional overhead or sum rate loss in comparison with the original SUCRe protocol \cite{sucre2017}.

\noindent\textit{\textbf{Step 3}: Contention Resolution and Pilot Repetition}. To resolve contentions distributively and uncoordinately, the $k$-th UE now has $\widehat{\alpha}_{t,k}$, which is the summation of the contending UEs signal gains with its own $\rho_k \sum\limits_{m\in \mathcal{V}_k}\beta^{(m)}_k \tau_p$.  However, the number of contenders $|\mathcal{S}_t|$ as well as the VRs of each UE are unknown by the users, leading to the only possibility of comparing its own overall gain with $\widehat{\alpha}_{t,k}$,  by computing $\frac{\rho_k}{\widehat{\alpha}_{t,k}} \sum\limits_{m\in \mathcal{V}_k}\beta^{(m)}_k \tau_p$. Hence, UEs using the SUCRe-XL protocol apply the following {\it decision rule}:
\begin{align}\label{eq: step3: decisionRule1}
\mathcal{R}_k:& \quad \sum_{m\in\mathcal{V}_k} \rho_k \beta^{(m)}_k \tau_p >  \widehat{\alpha}_{t,k}/2 + \epsilon_k \quad ({\rm repeat}),\\
\mathcal{I}_k:& \quad \sum_{m\in\mathcal{V}_k} \rho_k \beta^{(m)}_k \tau_p \leq  \widehat{\alpha}_{t,k}/2 + \epsilon_k \quad ({\rm inactive}).\label{eq: step3: decisionRule2}
\end{align}  
In this decision rule, the bias term $\epsilon_k$ is given by
\begin{equation}\label{eq: biasTerm}
\epsilon_k = \frac{\delta}{\sqrt{M_b}}\sum_{b\in\mathcal{V}_k} \beta^{(b)}_k
\end{equation}
where $\delta$ is an adjustable scale factor for finding a suitable operation point. As in \cite{sucre2017}, we adopt a $\delta = -1$.

There are four possible cases in a contention process: \textbf{\textit{i}}. Non-overlapping UEs win (false positive). Ex.: from Fig. \ref{fig:pilots} users 1 and 3 win. \textbf{\textit{ii}}. Only one UE wins. {\bf \textit{iii}}. None of the UEs win (false negative).  \textbf{\textit{iv}}.  Overlapping UEs win (false positive). Ex.:  from Fig. \ref{fig:pilots}, users 1 and 2 or 2 and 3 win.  Although case 1 is a false positive, there is no pilot collision. Therefore, cases 1 and 2 are successful attempts and there is the allocation of the RA pilot.  Case 4 is considered a pilot collision; {\emph{i.e.}, a pilot collision occurs if more than one UE in $\mathcal{S}_t$ retransmit in step 3 and have overlapping VRs. 

\noindent\textit{{\bf Step 4}: Allocation of Dedicated Payload Pilots} After the BS receives the repeated UL pilot transmissions from step 3, it tries to decode the message with new channel estimates from the repeated pilots. If the decoding goes well, the BS can allocate pilot sequences in the payload data blocks to the non-overlapping contention winners, followed by a replying DL message informing the successful connection and, if necessary, more information. If the decoding fails, the protocol failed to resolve that collision and the unsuccessful UE is instructed to try again after a random interval.  

\noindent\textit{\textbf{SUCRe-XL Complexity}} is equivalent to that of conventional SUCRe protocol. Although the computation of the precoding vector increases marginally at the BS with the number of SAs $B$, due to the sum of all different estimated channels in \eqref{eq:step2: DL pilot response}, the same precoding vector is used for all SAs, different than the precoding in \eqref{eq:step2: DL pilot response2} for the original SUCRe. While the original SUCRe has to compute $B$ different vector inner products in \eqref{eq:step2: DL pilot response2}, the proposed SUCRe-XL protocol has to compute a sum of $B$ vectors followed by a single vector inner product in \eqref{eq:step2: DL pilot response}. Also, each UE $k \in \mathcal{S}_t, \forall t$ has to estimate the sum of its large scale fading coefficients in SUCRe-XL protocol, which can be evaluated as the average received power of a beacon signal in a step 0, similarly as assumed in \cite{sucre2017}.

\section{Numerical Results} \label{sec:results}
It is assumed a 100 meter ULA with $M = 500$ antennas in a $200 \times 200$ m$^2$ square cell with $K = 1000$ uniformly distributed iUEs (crowded scenario) as illustrated in Fig. \ref{fig:array}, each user wants to become active with probability $P_a = 0.01$. It is considered $\tau_p = 10$ pilots, and transmit powers $\rho_k=q = 1 W, \, \forall k$.  Two channel models were deployed: \textit{\textbf{i}}) uncorrelated Rayleigh fading, as in eq. \eqref{eq: uncorr Rayleigh channel}, with $\mathbf{R}^{(b)}_{k} = \mathbf{I}_{M_b}$; \textit{\textbf{ii}}) correlated Rayleigh fading model, following eq. \eqref{eq:Rcorrel}, with $r=0.7$.

A baseline ALOHA-like performance has been included for comparison purpose, which treats pilot collision by retransmission after a random waiting time period, hence, contending users retransmit their pilots at random if collision occurs.

\noindent The  \textbf{\textit{Probability to Resolve Collision}} (PRC) is calculated numerically taking all resolved collisions per total number of collisions occurred. Simulations were carried out in sequential RA blocks fashion, where iUEs try to access the channel in each iteration. For each parameter value of the \textit{x}-axis ($B$ or P$_b$ in Fig. \ref{fig3}), it is simulated $10^4$ sequential RA blocks. If an attempt fails, UE makes another attempt with probability 0.5 in the subsequent blocks. It is given a limit of 10 RA attempts per UE, after which a failed access attempt is declared.

Fig. \ref{fig3}(a) depicts the PRC and the normalized mean square error (NMSE), given by $\mathbb{E}\{|\widehat{\alpha}_{t,k}-\alpha_t|^2 \}/\alpha_t$. It shows that increasing the number of  {SAs} $B$, which means reducing the number of antennas per {SA} $M_b$, since $M/B = M_b$, causes a progressive discrepancy on $\alpha_t$ estimation due to \eqref{eq: channel Hardening 1} and  \eqref{eq: favorable propagation between UEs} do not hold when $M_b$ decreases. Indeed, NMSE levels for the SUCRe-XL protocol deteriorate steadily when $B > 50$ for both channel models. To simplify this simulation, $|\mathcal{V}_k| = B, \; \forall k$.  The PRC starts increasing until $B = 25$ for the uncorrelated Rayleigh fading, and presents an optimal PRC value when $B = 10$ for the correlated Rayleigh fading model\footnote{The initial PRC increase is due to the SUCRe-XL decision rule associated with the possibility of users retransmitting the same RA pilot having non overlapping VRs, but then the reduced number of antennas per SA diminishes the channel hardening and favourable propagation effects, as well as the quality of the $\widehat{\alpha}_{t,k}$ estimates and, consequently, the PRC. Channel correlation highlights this effect, making the PRC starts to decrease with a lower $B$ value.}. $B=1$ corresponds to a spatial stationary regime.

\begin{figure}[htbp!]
\vspace{-3mm}
\centering
\includegraphics[trim={4mm 1mm 0mm 6mm},clip,width=.9\linewidth]{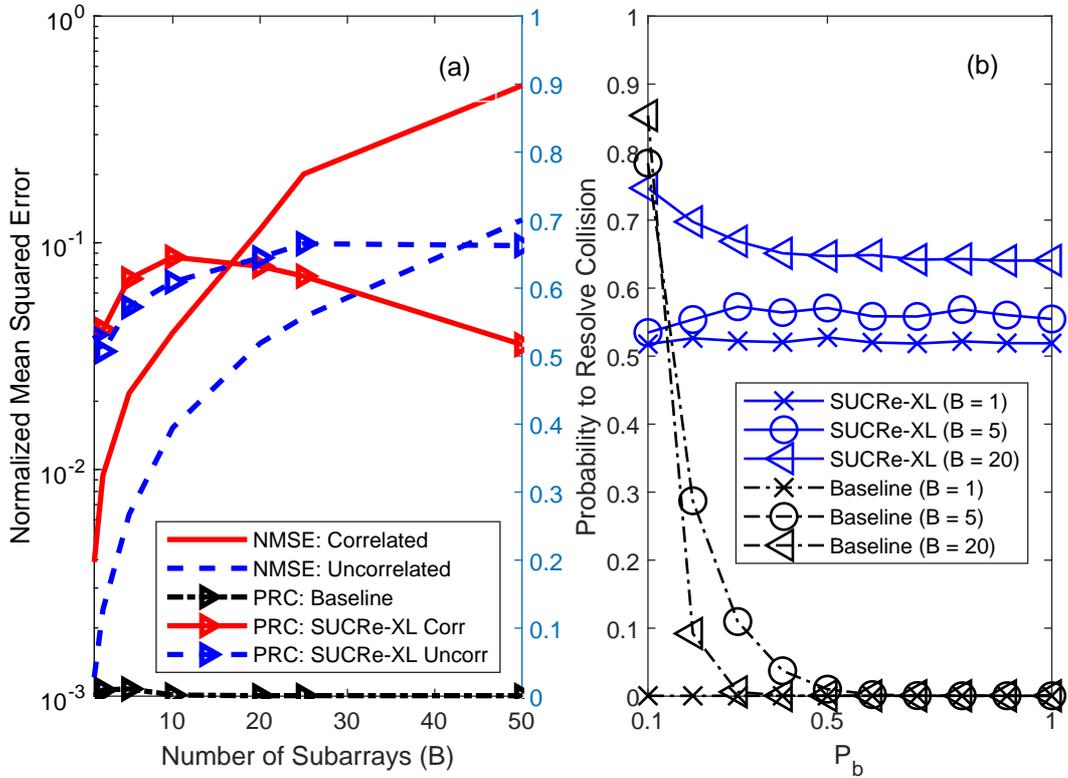}
\vspace{-3mm}
\caption{(a) \textit{Left y-axis}: NMSE to verify the $\alpha_t$ estimation. \textit{Right y-axis}: Probability to resolve collision (PRC) against the number of subarrays $B$. Baseline performance remains the same for both channel models. (b) PRC {\it vs.} $P_b$ for the uncorrelated Rayleigh model.}
\label{fig3}
\vspace{-1mm}
\end{figure}

Fig. \ref{fig3}(b) depicts the PRC for different probabilities of each SA being visible for a given UE, $P_b$. Notice that $P_b$ is
inversely proportional to the density of obstacles affecting transmitted signals. The probability of the VRs of $|\mathcal{S}_t|$ UEs in \eqref{eq: binomial distribution} not overlapping, given by $P_{\rm no} = ( (1-P_b)^{|\mathcal{S}_t|} + |\mathcal{S}_t| P_b (1-P_b)^{|\mathcal{S}_t|-1} )^B$, decreases with increasing $B$ and/or $P_b$. Thus, decreasing $P_b$ causes $|\mathcal{V}_k| \; \forall \; k$ to diminish at the BS side. Hence, the probability of pilot collisions in overlapping areas reduces when probability $P_b$ decreases, improving the SUCRe-XL PRC for $B = 20$, as well for the Baseline performance.  However, this could not be seen for the SUCRe-XL for $B = 5$, since the effect of the imposed constraint\footnote{To avoid  the  possibility  of  a  given  user  do  not see any subarray, while the average number of visible subarrays per user follows $\mathbb{E}[|\mathcal{V}_k|] = B \cdot P_b$.} $|\mathcal{V}_k|\geq 1$ is more noticeable when $P_b\leq 1/B$. Thus, when decreasing $P_b$ below the threshold $1/B$, the expected value of visible subarrays would decrease below 1, in such a way that the additional constraint $|\mathcal{V}_k| \geq 1$ turns to intervene more frequently, breaking the trend of the presented result in increasing the PRC with the decrease of $P_b$, as expected according to the $P_{\rm no}$ expression. Besides, this does not occur for $B = 1$ (stationary case), since the only SA existent is the entire linear array. Furthermore, the Baseline success probability grows abruptly comparing with the proposed protocol with $P_b$ reduction. This behavior might come from non-overlapping cases, when the decision rule would be unnecessary: the Baseline recognizes non-overlapping pilot collisions as successful attempts, while UEs in the SUCRe-XL protocol still have to decide to repeat the RA pilot, even when they are not overlapping. 

\noindent\textbf{\textit{Average Number of Access Attempts}}.
Numerical results in Fig. \ref{fig:avgNumberAccess}(a) shows the average number of RA attempts as a function of the number of iUEs. The fraction of UEs that could not access the network, {\it i.e.}, the portion that is unsuccessful in the maximum number of 10 RA attempts, is illustrated in Fig. \ref{fig:avgNumberAccess}(b). There is a clear advantage of SUCRe-XL in reducing the failed access attempts when exploiting the channel non-stationarities, supporting a higher number of UEs.

 \begin{figure}[!htbp]
 \vspace{-2mm}
 \centering
\includegraphics[trim={6mm 1mm 5mm 6mm},clip,width=0.9\linewidth]{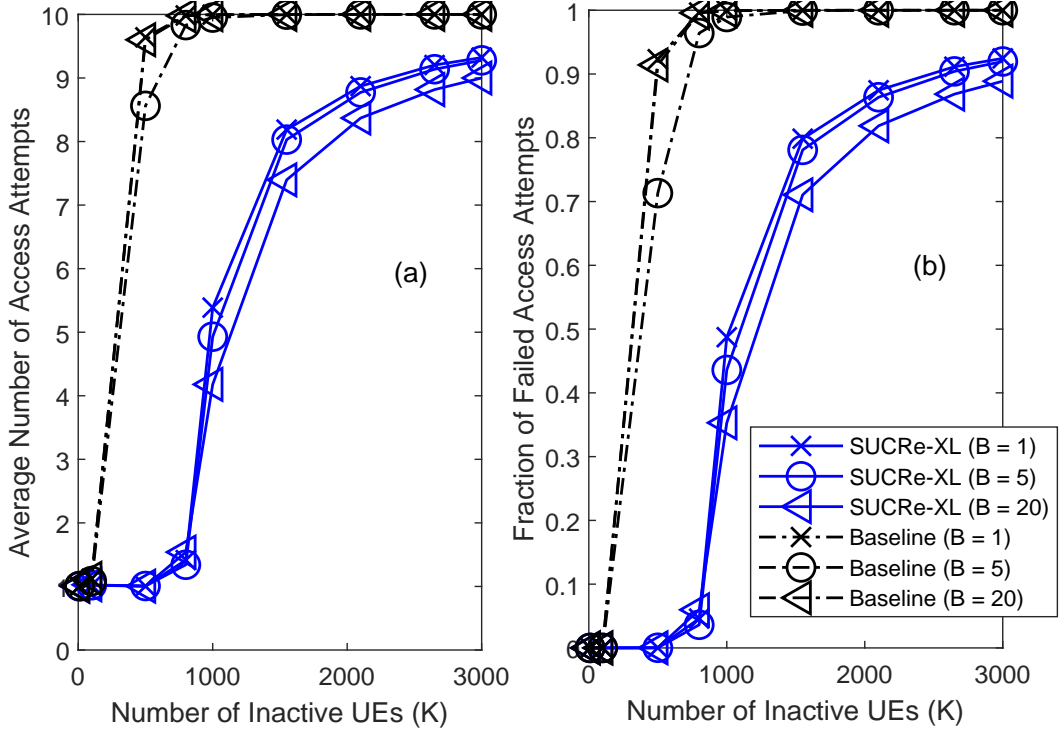}
\vspace{-2mm}
\caption{SUCRe-XL and Baseline performance in the uncorrelated Rayleigh fading model ($P_b = 0.5$). (a) Average number of RA attempts. (b) Probability of failed access attempts.}
 \label{fig:avgNumberAccess}
 \end{figure}
 
\begin{figure}[!htbp]
 \vspace{-6mm}
 \centering
\includegraphics[trim={6mm 6mm 5mm 4mm},clip,width=0.9\linewidth]{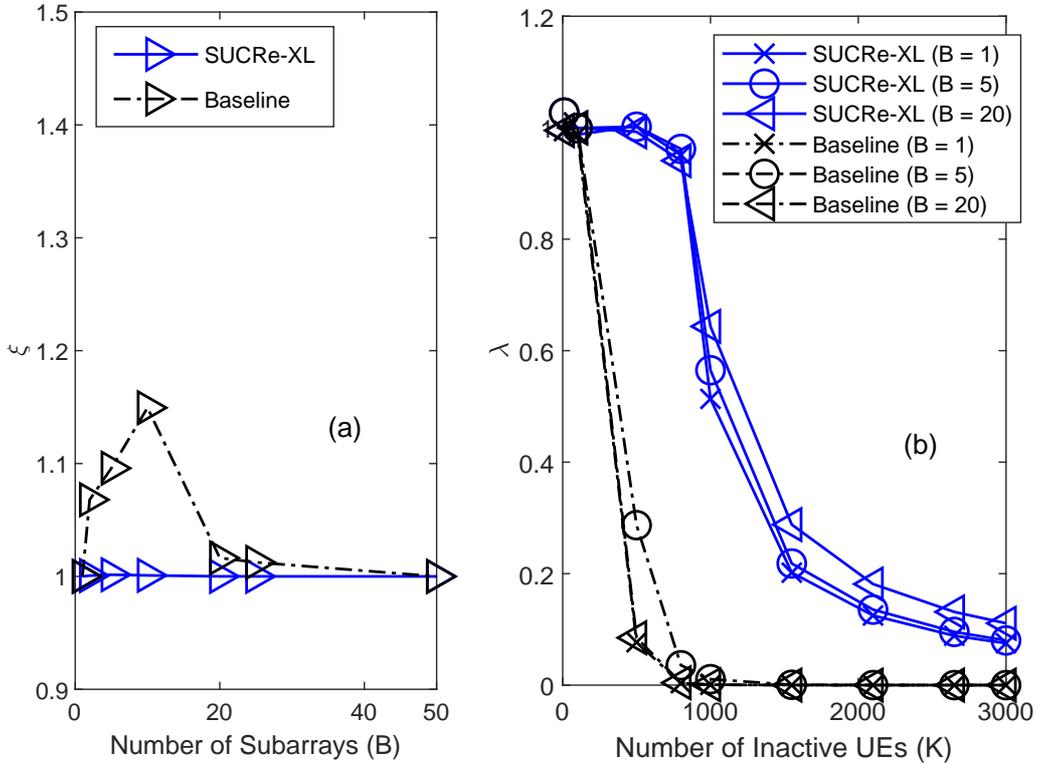}
\vspace{-3mm}
\caption{SUCRe-XL and Baseline performance in the uncorrelated Rayleigh fading model. (a) Average number of UEs per resolved collision {\it vs.} B. (b) Normalized number of accepted UEs for different numbers of iUEs, considering $10^4$ RA blocks.}
\label{fig: acceptedUEs}
\end{figure}

 Fig. \ref{fig: acceptedUEs}(a) depicts the {\it average number of accepted UEs per resolved collision} ($\xi$), showing that $\xi$ remains around one with increasing number of subarrays. Although $\xi$ is slightly higher for the Baseline scheme, the resolved collisions are much rarer in this simple scheme, as in Fig. \ref{fig3}(a). In the same scenario, Fig. \ref{fig: acceptedUEs}(b) indicates the {\it normalized number of accepted UEs} ($\lambda$) that realized successful attempts. Hence, in average, the total number of admitted UEs along the  $10^4$ RA blocks is given by $\Lambda= \lambda\cdot K \cdot P_a \cdot 10^4$. Indeed, non-stationary cases surpasses the stationary one, specially in (over)crowded mMTC scenarios, being able to manage a greater number of UEs.  

\section{Conclusion}\label{sec:conclusion}
Grant-based RA operating under massive antennas has demonstrated satisfactory performance  to handle multiple access attempts under (over)crowded scenarios, typically present in cMMB. Furthermore, XL-MIMO is a promising concept to surpass the performance of classical antenna structures. Hence, to take advantage of channel non-stationarities, an adapted SUCRe protocol for XL-MIMO has been proposed and compared.  Besides, the proposed protocol can support a higher number of active UEs, since it attains a reduced fraction of failed access attempts and reduces access latency}.
\appendix
\section{Proof of equation \ref{eq:step1: asymptoticPropagation of y}}
For simplicity, let $\rho_i$ be the same for all $i$; then we have:
\begin{align}
\left(\sum_{b\in \mathcal{M}} \mathbf{y}^{(b)}_t\right)^H = \sum_{b\in \mathcal{M}}\sum_{i \in S_t} \sqrt{\rho_i \tau_p} \mathbf{h}^{(b)H}_i + \mathbf{n}^{(b)H}_t.
\end{align}

Then, $$\left\lVert \sum_{b\in \mathcal{M}}\mathbf{y}^{(b)}_t \right\rVert^2  =  \left(\sum_{b\in \mathcal{M}} \mathbf{y}^{(b)}_t\right)^H\cdot\left(\sum_{b\in \mathcal{M}} \mathbf{y}^{(b)}_t\right) = \rho_i  \tau_p \sum_{b \in \mathcal{M}} \sum_{i\in S_t} \lVert \mathbf{h}_i^{(b)} \rVert^2 + 2 \rho_i \tau_p \sum_{b\in \mathcal{M}} \; \sum_{\substack{i,j \in S_t\\ i \neq j}} \mathbf{h}_{i}^{(b)H} \mathbf{h}_{j}^{(b)} +$$
\scalemath{1}{
\begin{aligned}
& + 2 \rho_i \tau_p \sum_{\substack{m, b \in \mathcal{M} \\ m \neq b}} \; \sum_{i, j \in S_t} \mathbf{h}_i^{(m)H} \mathbf{h}_j^{(b)} + 2\sqrt{\rho_i \tau_p} \sum_{m,b \in \mathcal{M}} \sum_{i \in S_t} \mathbf{h}_i^{(m)H} \mathbf{n}_t^{(b)}   + 2\sum_{\substack{m, b \in \mathcal{M} \\ m \neq b}} \mathbf{n}_t^{(m)H} \mathbf{n}_t^{(b)} + \sum_{b \in \mathcal{M}} \lVert \mathbf{n}_t^{(b)} \rVert^2
\end{aligned}
}

\noindent Dividing $\left\lVert \sum_{b\in \mathcal{M}}\mathbf{y}^{(b)}_t \right\rVert^2 $ by $M_b \rightarrow \infty$, components with different indices, as the second to the fifth, become zero, following property in eq. (\ref{eq: favorable propagation between UEs}). 
Furthermore, the first component obeys approximation (\ref{eq: channel Hardening 1}), resulting in $\beta_i$. The last term becomes noise variance, validating approximation (\ref{eq:step1: asymptoticPropagation of y}).  



\end{document}